\documentclass[mathleft]{an}
\usepackage{graphicx}
\usepackage{times}
\begin{document}

\Pagespan{1}{}
\Yearpublication{2007}%
\Yearsubmission{2007}%
\Month{11}%
\Volume{999}%
\Issue{88}%

\title{Modelling the time variation
 of the surface differential rotation 
in AB~Doradus and LQ~Hydrae}

\author{A.~F.~Lanza\thanks{Corresponding author:
  \email{nuccio.lanza@oact.inaf.it}\newline}
}
\titlerunning{Time variation of differential rotation in AB~Dor and LQ~Hya}
\authorrunning{A.~F.~Lanza}
\institute{
INAF-Osservatorio Astrofisico di Catania, Via S.~Sofia, 78, 95123 Catania, Italy
}

\received{29 August 2007}
\accepted{}
\publonline{later}


\abstract{Sequences of Doppler images of the young, rapidly rotating
late-type stars AB Dor and LQ Hya show that their equatorial
angular velocity and the amplitude of their surface differential
rotation vary versus time. Such variations can be modelled to
obtain information on the intensity of the azimuthal magnetic stresses
within stellar convection zones. 
We introduce a simple model in the framework of the mean-field theory and discuss briefly 
the results of its application to those  solar-like stars. 
  }
\maketitle

\section{Introduction}

\noindent
Doppler imaging techniques allow us to measure the surface 
differential rotation in rapidly rotating late-type stars by tracking the
longitudinal motion of starspots located at different latitudes 
(Collier Cameron 2007). 
Spe\-ci\-fi\-cal\-ly, the surface angular velocity $\Omega$ at colatitude $\theta$ is assumed to be given by
a solar-like law: 
\begin{equation}
\Omega(\theta) = \Omega_{\rm eq} - d\Omega \cos^{2} \theta, 
\end{equation} 
where 
$\Omega_{\rm eq}$ is the equatorial angular velocity and  $d\Omega$ is the pole-equator angular velocity difference. 
$\Omega_{\rm eq}$ and $d\Omega$ can be measured by fitting the shear of starspots along
 sequences of photospheric images covering successive rotations.
Alternatively, $\Omega$ and $d\Omega$ can be included as free parameters in a code that 
reproduces the line profile distortions due to starspots, thus obtaining their values by
a suitable $\chi^{2}$ minimization when a sufficiently long time series of line profiles is available. 

The long-term monitoring of surface differential rotation of the 
two late-type dwarfs AB Doradus and LQ~Hydrae shows that their
equatorial angular velocity and surface shear are functions of the time 
(see Donati et al. 2003a; Jeffers et al. 2007). 
It is interesting to note that the variations of $\Omega_{\rm eq}$ and $d\Omega$ are compatible with an internal angular
velocity uniform on cylindrical surfaces co-axial with the rotation axis (Donati et al. 2003a). 

\section{Modelling differential rotation variations}

Differential rotation is the result of the  re-distribution of 
angular momentum in a stellar convection zone under the action of  
meridional circulation, Reynolds stresses (i.e., the correlations of the components of the turbulent velocity in a
rotating star), and  torque by the Lorentz force (quantified by the azimuthal Maxwell stresses).  
The shear associated with differential rotation is opposed by the  turbulent
viscosity present in the convection zone that tends to make the star rotate rigidly. 

In order to model the time variation of the differential rotation, let us consider 
the equation for the angular momentum conservation in an inertial reference frame 
in the framework of the mean-field theory 
(e.g., R\"udiger \& Hollerbach 2004):
\begin{equation}
\frac{\partial}{\partial t} (\rho r^{2} \sin^{2} \theta \, \Omega) + \nabla \cdot {\vec \Theta} = 0,  
\label{angmomeq}
\end{equation}
where $r$ is the distance from the centre of the star,  $t$ the time, $\rho$  the density, 
$\Omega (r, \theta, t)$ the angular velocity and ${\vec \Theta}$ 
the  angular momentum flux vector given by:
\begin{eqnarray}
\lefteqn{{\vec \Theta} = (\rho r^{2} \sin^{2} \theta \, \Omega) {\vec u}_{\rm (m)} } \nonumber \\
 & & + \, r \sin \theta \langle \rho {\vec u}^{\prime} u_{\varphi}^{\prime} \rangle  
- \frac{r \sin \theta}{\tilde{\mu}} ({\vec B} B_{\varphi} + \langle {\vec B}^{\prime} B_{\varphi}^{\prime} \rangle),
\label{theta_expr}
\end{eqnarray}
where ${\vec u}_{\rm m}$ is the meridional circulation, ${\vec u}^{\prime}$  the fluctuating velocity, 
$\vec B$ the mean magnetic field, ${\vec B}^{\prime}$ the fluctuating \linebreak 
magnetic field 
and $\tilde{\mu}$ the magnetic permeability;  
angular brackets $\langle \rangle$ indicate the Reynolds mean, 
defining the mean quantities. Note that in Eq.~(\ref{theta_expr}) the first term 
specifies  the angular momentum transport by  meridional circulation, the second by Reynolds stresses and the
third by Maxwell \linebreak 
stresses, respectively. 

The Reynolds stresses can be written as: 
\begin{equation}
\langle \rho u^{\prime}_{i} u^{\prime}_{j} \rangle = 
-\eta_{\rm t} \left( \frac{\partial u_{i}}{\partial x_{j}} +
\frac{\partial u_{j}}{\partial x_{i}} \right)
 + \Lambda_{ij},
\label{Rey_stress}
\end{equation}
where the first term represents the effect of the turbulent viscosity and the second one is the so-called 
$\Lambda$-effect
due to the correlation of the fluctuating velocity components produced by the action of the Coriolis force in a
rotating star (R\"udiger 1989). 

\begin{figure}[t]
\includegraphics[width=8cm,height=6cm]{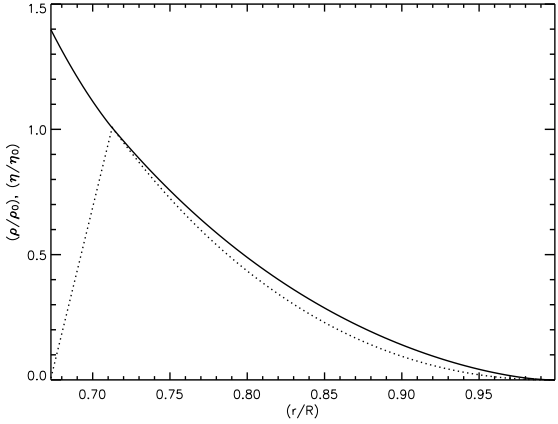}
\caption{The dynamic turbulent viscosity, computed according to 
standard mixing-length theory (dotted line) and the density (solid line), 
normalized to their respective values at the base of the convection zone,
in the model for a solar-like star adopted by Lanza (2007). The convection zone extends from 0.713 $R$ to the surface,
whereas an overshoot layer is assumed to extend between 0.673 and 0.713 $R$ with a linearly increasing
$\eta_{\rm t}$. }
\label{turb_visc}
\end{figure}

The conservation of the total angular momentum of the convection zone leads to the following 
boundary conditions: 
\begin{equation}
\Theta_{r} = 0 \mbox{ for $r=r_{\rm c}, \, R$,}
\label{bcond}
\end{equation}
where $r_{\rm c}$ is the radius at the lower boundary of the convection zone and $R$ is the radius of the star. 
We separate the contribution of the turbulent viscosity stress tensor from the other terms in Eq.~(\ref{theta_expr})
and adopt a dynamic turbulent viscosity 
that depends only on the radial co-ordinate, i.e.,  
 $\eta_{\rm t} = \eta_{\rm t}(r)$ (see Fig.~\ref{turb_visc} for details). 
  In such a way, the equation for the angular velocity derived from Eqs.~(\ref{angmomeq}), (\ref{theta_expr}) and
(\ref{Rey_stress}) becomes:
\begin{eqnarray}
\lefteqn{\frac{\partial \Omega}{\partial t} - \frac{1}{\rho r^{4}} \frac{\partial}{\partial r} 
\left(  r^{4} \eta_{t} \frac{\partial \Omega}{\partial r} \right) }   \nonumber \\
 &  & - \, \frac{\eta_{t}}{\rho r^{2}}
\frac{1}{(1- \mu^{2})} \frac{\partial}{\partial \mu} \left[ (1 - \mu^{2})^{2} \frac{\partial \Omega}{\partial \mu}\right] = S,  
\label{angvel}
\end{eqnarray} 
where $\mu \equiv \cos \theta$ and the source term $S$ is given by:
\begin{equation}
S = - \frac{\nabla \cdot {\vec \tau}}{\rho r^{2} (1- \mu^{2})}, 
\label{sourceangvel}
\end{equation}
${\vec \tau}$ being a vector whose components are: 
\begin{eqnarray}
\lefteqn{\tau_{i} =  \rho r^{2} \sin^{2} \theta \Omega  u_{\rm (m) i} } \nonumber \\
& & + \, r \sin \theta \left[ \Lambda_{i \varphi} - \frac{1}{\tilde{\mu}} \left( B_{i} B_{\varphi} + 
 \langle B_{i}^{\prime} B_{\varphi}^{\prime} \rangle \right) \right].  
\label{tau}
\end{eqnarray}
The boundary conditions given by Eq.~(\ref{bcond}) can be recast as:
\begin{equation}
\frac{\partial \Omega}{\partial r} = 0 \mbox{ for $r=r_{\rm c}, \, R$.}
\label{bcond1}
\end{equation}
The solution of the angular velocity equation (\ref{angvel}) with the boundary conditions
(\ref{bcond1}) can be expressed in terms of an appropriate { Green function}
(see Lanza 2007 for details):
\begin{eqnarray}
\Omega (r, \theta, t)  & = & \int_{V} [\nabla \cdot {\vec \tau (r^{\prime}, \theta^{\prime}, t)}] G (r, \theta, r^{\prime}, \theta^{\prime}) dV^{\prime} = \nonumber \\
& = & \int_{V} ({\vec \tau} \cdot \nabla^{\prime} G) dV^{\prime},
\label{green}
\end{eqnarray}
where the integration with respect to the variables $r^{\prime}$ and $\theta^{\prime}$ is extended 
to the volume of the convection zone $V$ and $\nabla^{\prime} G$ is the gradient of $G$ with respect to the variables
$r^{\prime}$ and $\theta^{\prime}$. The second equality in Eq.~(\ref{green}) follows by 
the application of Gauss's Theorem and the vanishing of $\tau_{\rm r}$ at the boundaries of the convection zone. 
 Note that Eq.~(\ref{green}) is valid when the timescale of variation of 
$\vec \tau$ is much longer than the diffusion timescale of the convection zone $t_{\rm diff} \simeq (R-r_{\rm c})^2 / \nu_{\rm t}$, with
$\nu_{\rm t} \propto \eta_{\rm t}/\rho$ the average kinematic turbulent viscosity of the convection zone. 
If this assumption is not valid, the solution can still be expressed as a linear combination of  
terms each depending on an appropriate Green function (see Lanza 2007). 
For the sake of simplicity, however, we prefer to work with Eq.~(\ref{green}), so we retain this assumption.  

Note that the Green function $G$ depends on $\eta_{\rm t}$ and $\rho$, but does not depend on the 
angular momentum transport terms.

We shall obtain information on the averaged intensity of the angular momentum transport 
vector $\vec \tau$ in the framework of two different hypotheses, i.e.: 
a) the internal angular velocity is uniform on cylindrical surfaces co-axial with the rotation axis 
(Taylor-Proudman regime); or,  
b) the torque is localized between, say, $r_{1} \leq r \leq r_{2}$ within the convection zone.

\section{Applications and results}

Let us consider first the case of a stellar convection zone rotating in the Taylor-Proudman regime. 
This is probably the case for { rapid rotators}, like AB Dor and LQ Hya.  
In such a case, Eq.~(\ref{green}) can be written in the form:
\begin{equation}
\Delta \Omega (s, t) = \int_{0}^{R} {\cal{C}}(s^{\prime}) \tau_{\rm s}(s^{\prime}, t) \, 
\frac{\partial G(s, s^{\prime})}{\partial s^{\prime}} \, d s^{\prime}, 
\label{omegas}
\end{equation}
where $\Delta \Omega$ is the variation of the angular velocity with respect to a reference rotation
state (e.g., that with ${\vec \tau} = 0$),  $s$  is the distance from the stellar rotation axis 
in a cylindrical polar reference frame and ${\cal{C}}(s)$ the lateral area of the cylindrical surface 
of radius $s$ within the convection zone. 
If we pose: 
\begin{equation}
M = \max \left\{\left| \frac{\partial G}{\partial s^{\prime}} \right| \right\} \mbox{ in the convection zone,}
\end{equation}
then, Eq.~(\ref{omegas}) can be applied to obtain the inequality (see Lanza 2006a for details):
\begin{equation}
\int_{V} | \tau_{s} | dV^{\prime} \geq \frac{|\Delta \Omega|}{M}. 
\label{tauestim}
\end{equation}
Equation~(\ref{tauestim}) provides us with a lower limit for the average radial angular momentum transport term
in the convection zone. 
Under the additional hypotheses that the reference state corresponds to solid-body rotation with the total angular momentum of the star, and 
${\vec \tau} = -(s/\tilde{\mu}) B_{\varphi} {\vec B} $, i.e., 
only the azimuthal mean-field Maxwell stresses contribute to the deviation 
of the differential rotation from the reference state (cf. Covas, Moss \& Tavakol 2005), 
we find:  
\begin{eqnarray}
\mbox{min} \left\{ |B_{s} B_{\varphi}| \right\} = 8.22 \times 10^{-4} \left( \frac{\alpha_{\rm ML}}{1.75}\right)^{4/3}  \nonumber \\
\times \left( \frac{M}{M_{\odot}}\right)^{2/3} 
 \left( \frac{L}{L_{\odot}}\right)^{1/3}
\left( \frac{R}{R_{\odot}}\right)^{-5/3} |d \Omega| \mbox{ ~~~T$^{2}$},
\label{scalebb}
\end{eqnarray}
and the ratio of the dissipated power to the stellar luminosity (Lanza 2006a): 
\begin{eqnarray}
\frac{P_{\rm diss}}{L} = 4.40 \times 10^{-5} \left( \frac{\alpha_{\rm ML}}{1.75}\right)^{4/3} 
\left( \frac{M}{M_{\odot}}\right)^{2/3}  \nonumber \\
\times \left( \frac{L}{L_{\odot}}\right)^{-2/3} 
\left( \frac{R}{R_{\odot}}\right)^{4/3} (d\Omega)^{2};
\label{scalepdiss}
\end{eqnarray}
where the surface shear $d \Omega$  is measured in mrad d$^{-1}$, $\alpha_{\rm ML}$ is the ratio of the mixing-length to the pressure scale-height, $M$ the mass, $R$ the radius and $L$ the luminosity of the star. 
\begin{table}[!t]
 \centering
 \begin{tabular}{@{}crcccc@{}}
  \hline
 & & & & & \\
Star &  {$d\Omega$~~~}  & min $\left\{ |B_{s} B_{\varphi} |\right\}$ & $B_{\rm min}$ & $B_{\rm eq}$ 
 &  $P_{\rm diss}/L$ \\
 & & & & &   \\
 \hline
 & & & & & \\
AB Dor & 59.6 & 0.0385  &  0.196 & 0.475 & 0.23 \\
AB Dor & 96.7 & 0.0624  &  0.250 & 0.475 & 0.60\\
LQ Hya & -48.7 & 0.0336  &  0.183 & 0.495 & 0.15 \\
LQ Hya & 200.9 & 0.1386  & 0.372 & 0.495 & 2.52\\
 & & & &  & \\
\hline
\end{tabular}
\caption{$d\Omega $: surface shear  in mrad d$^{-1}$; $B_{\rm min}$: minimum magnetic field, obtained for $B_{s} = B_{\varphi}$; $B_{\rm eq}$: mean equipartition field. Magnetic field intensities are given in Teslas. }
\label{table1}
\end{table}
The results of the application of these equations to AB~Dor and LQ~Hya are listed in Table~\ref{table1},
for $\alpha_{\rm ML} = 1.75$ and adopted stellar parameters $M = 1.0$ M$_{\odot}$, $ R =1.05$ R$_{\odot}$, 
$L= 0.618$ L$_{\odot}$ for AB~Dor and $M=0.95$ M$_{\odot}$, $R=0.95$ R$_{\odot}$ and $L=0.506$ L$_{\odot}$
for LQ~Hya, respectively. The first column in Table~\ref{table1} 
indicates the name of the star, the second the observed surface shear, as obtained from 
Zeeman Doppler Imaging based on Stokes V data (see  Donati et al. 2003a), 
the third the minimum Maxwell stress, the fourth the minimum
mean field obtained for $B_{\rm s} = B_{\varphi}$, the fifth the mean equipartition field derived from the 
average kinetic energy of convective motions computed according to the mixing-length theory, and the last
the ratio of the dissipated power to the stellar luminosity. 
The minimum field strengths turn out to be smaller than the average equipartition values.
However, 
Zeeman Doppler Imaging provides us with an estimate of the radial mean field 
$B_{\rm s} \approx 0.01$ T in AB~Dor and LQ~Hya (Donati et al. 2003b), 
implying that the toroidal mean field is significantly stronger than the 
minimum values listed in Table~\ref{table1}, i.e., $B_{\varphi} \approx 3-14$ T. Such strong toroidal 
fields produce a radial Lorentz force
that opposes the centrifugal force in a rotating star leading to a perturbation of its oblateness 
and therefore of the quadrupole
moment of its outer gravitational potential. If the active star is a member of a close binary system, 
this leads to a variation of the orbital period through a mechanism originally proposed by 
Matese \& Whitmire (1983) and further discussed by Lanza (2005) and references therein.  

Note that the power dissipated in the turbulent convection zone of LQ~Hya exceeds the stellar luminosity
in the case of the largest observed shear, i.e., $d \Omega = 200.9$ mrad d$^{-1}$.  
This does not pose  a serious problem 
if the duration of such episodes of large shear is short in comparison to 
the phases with nearly rigid rotation
because the large energy reservoir represented by the thermal content of the stellar convection zone will average out 
the energy loss 
over a timescale comparable with the Kelvin-Helmoltz timescale of the envelope. Moreover, the  
uncertainty in the determination of the largest $d\Omega$ excursion of LQ~Hya 
(cf. Donati et al. 2003a) makes the estimate of the corresponding energy dissipation rate largely uncertain. 

Let us now consider the case in which  magnetic stresses are localized close to the base of the convection zone, 
without assuming a specific internal rotation regime. 
In this case, an expression similar to Eq.~(\ref{tauestim}) can be obtained
(cf. Lanza 2007). Specifically, we assume that 
${\vec \tau} = - (r \sin \theta / \tilde{\mu}) {\vec B} B_{\varphi}$ is 
localized in the overshoot layer, i.e., \linebreak 
 $0.67 \leq (r/R) \leq 0.71$ and that the differential rotation observed
at the surface is equal to that at $r=0.99 R$.  

In the case of the largest shear observed in LQ Hya, this yields a minimum average 
Maxwell stress over the overshoot layer 
$\left\{B_{\rm r} B_{\varphi}\right\}_{\rm min} \simeq 0.063$~T$^{2}$, which gives 
fields of the same order of magnitude of those obtained above under the Taylor-Proudman hypothesis.

\section{Discussion and conclusions}

We modelled the observed time variation of the differential
rotation in AB~Dor and LQ~Hya under the hypotheses that the 
azimuthal Maxwell stresses rule the 
changes of their surface shear and the internal angular velocity 
depends only on the distance from the rotation axis (Taylor-Proudman regime).  
We obtained that the average intensity of the mean field Maxwell stres\-ses is 
$ |B_{s} B_{\varphi} |  \sim 0.03 - 0.14 $ T$^{2}$, 
implying azimuthal mean fields $B_{\varphi} \sim (3-10)$ T for $B_{\rm s} \sim 0.01$ T. 
Similar Maxwell stresses are obtained if the magnetic torque is assumed to be confined within 
the overshoot layer $0.67 \leq (r/R) \leq 0.71$ and no restrictions are imposed on the internal
rotation law.  

It is interesting to note that 
azimuthal magnetic fields of $3-10$ T, occupying a sizeable fraction of the convection zone, 
have been invoked to explain orbital period changes observed in late-type active binaries (Lanza, Rodon\`o \linebreak
\& Rosner 1998; Lanza \& Rodon\`o 2004; 
Lanza 2005, \linebreak
2006b). An $\alpha$-effect related to an 
instability of the magnetic field itself (e.g., magnetic buoyancy instability, Brandenburg \& Schmitt 1998;
 or 
magneto-rotational instability, R\"udiger et al. 2007) seems to be necessary to produce such super-equi\-par\-ti\-tion fields, 
possibly acting in combination with differential rotation.  
The energy dissipated by  turbulence, estimated according to  
standard mixing-length arguments, may
exceed stellar luminosity in the case of the largest surface shear observed in LQ Hya. However, 
the thermal equilibrium of the convection zone can be  significantly affected only if those large shear  
episodes 
last more than 
$\sim 10-20$ per cent of the time.  Note also that a
mixing-length estimate for the turbulent viscosity may not be 
appropriate for a rapidly rotating star (see Lanza 2006a for details).  

In the present work, we adopted a point of view analogous to that of Covas et al. (2005) who modelled the 
variation of stellar differential rotation considering only the \linebreak 
torque exerted by the Lorentz force 
and neglecting the roles of meridional flow and $\Lambda$-effect. As a matter of fact, it is difficult to
evaluate the perturbations of the meridional flow and of the Reynolds stresses produced by the magnetic field
because they depend critically on the approximations made in the treatment of stellar turbulence in a rotating 
star. Nevertheless,   alternative models for the variation of differential rotation have been investigated, 
such as those based on a time-dependent component of the meridional flow (Rempel 2006, 2007) or 
 the magnetic quenching of the $\Lambda$-effect
(R\"udiger et al. 1986). The present approach 
can be generalized to obtain amplitudes of the perturbations of the corresponding 
terms in Eq.~(\ref{tau}), but we shall not pursue this application here (see Lanza 2007).

Finally, it is interesting to note that some inference on the amplitude of
variation of the surface shear in the case of very active stars can be
obtained not only by means of Doppler imaging techniques, but also by  an appropriate ana\-ly\-sis of their 
long-term wide-band photometry  (e.g., \linebreak 
Rodon\`o et al. 2001; Messina \& Guinan 2003). 
For example, in the case of LQ~Hya,
it is worth comparing the Doppler imaging results by Donati et al. (2003a) with 
the photometrically determined seasonal rotation periods by 
Kov\'ari et al. (2004).

\acknowledgements
The author wishes to thank the organizers of the 5th Potsdam Thinkshop for a very topical and
stimulating \linebreak 
meeting. He is grateful to Prof. A.~Collier Cameron and \linebreak 
Prof. R.~Tavakol for  useful discussions. 
Active star research at INAF-Catania Astrophysical Observatory and the Department of Physics
and Astronomy of Catania University is funded by MIUR ({\it Ministero 
dell'Uni\-ver\-si\-t\`a e della Ricerca}),
and by {\it Regione \linebreak 
Siciliana}, whose financial support is gratefully
acknowledged.

This research has made use of the ADS-CDS databases, operated at the CDS, Strasbourg, France.

\appendix

\end{document}